\documentclass{aastex}
\usepackage{emulateapj5}
\usepackage{apjfonts}

\slugcomment{UTAP-587}

\shorttitle{Towards unravelling the structural distribution of UHECR sources}
\shortauthors{Takami \& Sato}

\begin{document}

\title{Towards unravelling the structural distribution of ultra-high-energy cosmic ray sources}

\author{Hajime Takami\altaffilmark{1} and Katsuhiko Sato\altaffilmark{1,2}}
\email{takami@utap.phys.s.u-tokyo.ac.jp}

\altaffiltext{1}{Department of Physics, School of Science, the University of Tokyo, 7-3-1 Hongo, Bunkyoku, Tokyo 113-0033, Japan}
\altaffiltext{2}{Research Center for the Early Universe, School of Science, the University of Tokyo, 7-3-1 Hongo, Bunkyoku, Tokyo 113-0033, Japan}

\begin{abstract}
We investigate the possibility that near future observations of 
ultra-high-energy cosmic rays (UHECRs) can unveil 
their local source distribution, which reflects the observed local structures 
if their origins are astrophysical objects. 
In order to discuss this possibility, 
we calculate the arrival distribution of UHE protons 
taking into account their propagation process in intergalactic space 
i.e. energy losses and deflections by extragalactic magnetic field (EGMF). 
For a realistic simulation, 
we construct and adopt a model of a structured EGMF 
and UHECR source distribution, 
which reproduce the local structures actually observed around the Milky Way. 
The arrival distribution is compared statistically 
to their source distribution using correlation coefficient. 
We specially find that UHECRs above $10^{19.8}$eV are best indicators 
to decipher their source distribution within 100 Mpc, 
and detection of about 500 events on all the sky 
allows us to unveil the local structure of UHE universe 
for plausible EGMF strength and the source number density. 
This number of events can be detected by 
five years observation by Pierre Auger Observatory. 
\end{abstract}

\keywords{cosmic rays --- methods: numerical --- IGM: magnetic fields ---
galaxies: general --- large-scale structure of the universe}

\section{Introduction} \label{intro}

The origin of ultra-high-energy cosmic rays (UHECRs) 
above $10^{19}$eV is one of the most intriguing problems 
in astroparticle physics. 
Akeno Giant Air Shower Array (AGASA) found 
statistically significant small-scale clusterings 
of observed UHECR events with large-scale isotropy \citep*{takeda99}. 
The AGASA data set of 57 events above $4 \times 10^{19}$eV 
contains four doublets and one triplet 
within separation angle of $2^{\circ}.5$, 
consistent with the experimental angular resolution. 
The chance probability of observing such multiplets 
under an isotropic distribution is only about 1\% \citep*{hayashida00}. 
A combination of the results of many UHECR experiments 
(including AGASA) also revealed eight doublets and two triplets 
within $4^{\circ}$ on a totally 92 events 
above $4 \times 10^{19}$eV \citep*{uchihori00}. 
These multiplets suggest that the origins of UHECRs are point-like sources. 
For identification of UHECR sources, 
arrival directions of UHECRs have been observed in detail 
by High Resolution Fly's Eye (HiRes) and 
Pierre Auger Observatory (Auger). 
However, so far, these experiments have reported no significant clustering 
on the arrival distribution above $4 \times 10^{19}$eV 
\citep*{abbasi05,mollerach07}. 

Recently, several classes of astrophysical objects 
in many literature has tested 
for positional correlations with observed arrival directions of UHECRs. 
The correlations with BL Lac objects were discussed 
on the assumptions of smaller deflection angles of UHECRs 
than the experimental angular resolution and/or neutral primaries 
\citep*{tinyakov01}, 
and in consideration with the deflection 
by Galactic magnetic field (GMF) \citep*{tinyakov02}. 
\cite{gorbunov05} considered various classes of 
powerful extragalactic sources for the correlation study 
including small corrections of UHECR arrival directions by GMF. 
\cite{hague07} discussed the correlation with nearby 
active galactic nuclei (AGNs) from RXTE catalog of AGNs. 
However, these studies have not taken into account 
UHECR propagation in extragalactic space. 
UHECRs above $8 \times 10^{19}$eV lose 
a significant fraction of their energies by photopion production in collision 
with the cosmic microwave background (CMB) photons 
during their propagation \citep*{berezinsky88,yoshida93}. 
Thus, UHECRs have {\it horizons}, which are the maximum distances 
of their sources that UHECRs can reach the Earth, 
even if their energies are below $8 \times 10^{19}$eV at the Earth. 
The positional correlations between arrival directions of UHECRs 
and their source candidates outside the horizons are not significant. 
(In \cite{hague07}, only nearby AGNs within the horizons are considered.) 
In addition to the UHECR horizons, 
deflections due to extragalactic magnetic field (EGMF) are also important 
since extragalactic cosmic rays are propagated for a much greater distance 
than in Galactic space. 
Propagation process of UHECRs should be considered 
in such correlation studies. 

\cite{yoshiguchi03} investigated the correlation 
between the arrival distribution of UHECRs and their source distribution 
taking into account UHECR propagation in intergalactic space 
with a uniform turbulent EGMF whose strength is 1 nG 
and coherent length is 1 Mpc. 
The authors adopted a source distribution 
with $10^{-6}~{\rm Mpc}^{-3}$ that reproduced the local structures 
and the AGASA results. 
They concluded that detection of 
a few thousand events above $4 \times 10^{19}$eV 
reveal observable correlation with the sources within 100 Mpc. 

However, a uniform turbulent field is not realistic EGMF model. 
Faraday rotation measurements indicate magnetic field strengths 
at the $\mu$G level within inner region ($\sim$ central Mpc) 
of galaxy clusters \citep*{kronberg94}. 
The evidence for synchrotron emission 
in numerous galaxy clusters \citep*{giovannini00} 
and in a few cases of filaments \citep*{kim89,bagchi02} 
also seems to suggest the presence of magnetic fields 
with $0.1-1.0\mu$G at cosmological structures. 
Several numerical simulations of large-scale structure formation 
have confirmed these magnetic structures \citep*{sigl03,dolag05}. 

Based on these studies, in recent years, 
we calculated propagation of UHE protons in a structured EGMF 
which well reproduces the local structures actually observed 
and simulated their arrival distributions 
with several normalizations of EGMF strength 
and several number density of UHECR sources \citep*{takami07}. 
We constrained the source number density 
to best reproduce the AGASA results. 
As a result, 
$10^{-5}~{\rm Mpc}^{-3}$ is the most appropriate number density, 
which is weakly dependent on EGMF strength. 
(In rectilinear propagation, similar number density is also obtained 
in \cite{blasi04,kachelriess05})
However, this has large uncertainty 
due to the small number of observed events at present. 
$10^{-4}~{\rm Mpc}^{-3}$ and $10^{-6}~{\rm Mpc}^{-3}$ 
are also statistically allowed. 
Therefore, it is useful to deliberate the correlation between 
the arrival distribution and the source distribution 
in the case of these number densities. 
Note that we revealed in the paper that 
this uncertainty will be solved by future increase of detected events. 

In this study, 
we calculate the arrival distribution of UHECRs, 
taking their propagation process into account, 
and investigate the correlation the arrival distribution 
and their source distribution in the future. 
A structured EGMF model and source distribution 
which can reproduce the local universe actually observed are adopted. 
The source number density and the EGMF strength are treated as parameters 
since these have some uncertainty. 
A goal of this study is that 
we understand the number of observed events to start to observe 
the UHECR source distribution by UHECRs 
and how much the correlation is expected in the future. 

Auger has already detected more events above $10^{19}$eV 
than those observed by AGASA \citep*{roth07}. 
Nevertheless, the event clustering has not observed, as mentioned above. 
It might be due to EGMF and/or 
GMF strong enough not to generate the multiplets or 
statistical fluctuation for small number of observed events 
at highest energies. 
In any case, we should predict and discuss 
how the arrival distribution reflects UHECR source distribution. 

Chemical composition of UHECRs is very important for the correlation. 
If UHECRs are heavier components, 
magnetic deflections are larger and the correlation is worse. 
One of observables for study of UHECR composition is 
the depth of shower maximum, $X_{\rm max}$, 
which can be measured by fluorescence detectors. 
Its average value $\left< X_{\rm max} \right>$ is dependent on 
UHECR composition and energy. 
HiRes reported that composition of cosmic rays above $10^{19}$eV 
is dominated by protons as a result of $X_{\rm max}$ measurement 
\citep*{abbasi05b}. 
Recent result by Auger is compatible to the HiRes result 
within systematic uncertainties \citep{unger07}. 
However, they concluded that 
the interpretation of $\left< X_{\rm max} \right>$ distribution 
is ambiguous because of the uncertainties of hadronic interaction 
at highest energies. 
Thus, UHECR composition at highest energies is controversial at present. 
Despite that, in this study, 
we assume that all UHECRs are protons 
since composition above $10^{19}$eV has proton-like feature.

This paper is organized as follows: 
in section \ref{model} we provide our models 
of UHECR source distribution and a structured EGMF. 
In section \ref{method} we explain our calculation method 
for the arrival distribution with UHECR propagation 
and statistical method. 
In section \ref{results}, 
The results of the correlation 
between the arrival distribution of UHE protons 
and their source distribution. 
We summarizes this study in section \ref{conclusion}.

\section{Source Distribution and Magnetic Field} \label{model}

In this section, 
our models of UHECR source distribution 
and a structured EGMF are briefly explained. 
These models are almost the same as those in our previous work. 
More detailed explanations are written in \cite{takami06}. 

These models are constructed from 
{\it Infrared Astronomical Satellite} Point Source Catalogue Redshift 
Survey ({\it IRAS} PSCz) catalog of galaxies \citep*{saunders00}. 
This is a catalog of flux-limited galaxy survey 
with large sky coverage ($\sim$ 84\% of all the sky). 
Thus, this is appropriate galaxy catalog for the purpose. 
The selection effects for observation are corrected 
with luminosity function of the {\it IRAS} catalog \citep*{takeuchi03}. 
We use a set of galaxies after the correction within 100 Mpc 
(called our sample galaxies below) for construction of our models 
since small number of galaxies can be observed above 100 Mpc. 
We adopt $\Omega_m=0.3,~\Omega_{\Lambda}=0.7$ and 
$H_0=71~{\rm km}~{\rm s}^{-1}~{\rm Mpc}^{-1}$ 
as the cosmological parameters. 

We assume that subsets of our sample galaxies 
with specific number densities are UHECR source distributions. 
$10^{-4},~10^{-5}$ and $10^{-6}~{\rm Mpc}^{-3}$ is considered 
as the source number density. 
For the source number densities, 
we randomly select galaxies from our sample galaxies 
with probabilities proportional to the absolute luminosity of each galaxy. 
This method allows constructing source distributions 
to reflect the cosmic structures. 
Source distribution above 100 Mpc is assumed to be isotropic 
and luminosity distribution of galaxies follows the luminosity function. 
We take the sources until 1 Gpc into account. 
All sources are assumed to be have the same power for injection 
of UHE protons. 

Our EGMF model also reproduces the local structures actually observed 
around the Milky way. 
Several simulations of cosmological structure formation with magnetic field 
have found that EGMF roughly traces baryon density distribution. 
\citep*{sigl03,dolag05}. 
Our structured EGMF model is constructed with simple assumptions 
in addition to these results. 
We constructed matter density distribution from our sample galaxies 
with a spatial resolution of $1 \rm{Mpc}$ equal to the 
correlation length of our EGMF model, $l_c$. 
In each cell, 
strength of the EGMF is related to matter density, $\rho$, as 
$\vert B_{\rm EGMF} \vert \propto {\rho}^{2/3}$ and 
a turbulent magnetic field with the Kolmogorov spectrum is assumed. 
The strength of EGMF is normalized to $B=0.0,~0.1$ and $0.4\mu$G at a cell 
that contains the center of the Virgo Cluster 
since EGMF strength has some uncertainty as mentioned in section 1. 
These three normalizations are used for investigation 
of the spatial correlation between the arrival distribution 
and the source distribution. 
Since the EGMF model and the source distribution are constructed 
from the same galaxy sample, 
UHECR sources are in the magnetized structure. 
Note that volume of about 95\% has no magnetic field in our EGMF model.

\section{Method of Calculation} \label{method}

\subsection{Calculation of arrival distribution of UHE protons} \label{method_calc}

Once a source distribution is given, 
the arrival distribution of UHE protons can be calculated 
by calculating propagation of protons from their sources to the Earth. 
In this section, 
we describe fundamental processes on proton propagation 
in intergalactic space 
and our calculation method of the arrival distribution. 

UHE protons lose their energies in collision 
with CMB during their propagation 
in intergalactic space \citep*{berezinsky88,yoshida93}. 
Higher energy photon backgrounds (e.g.infrared, optical, ultraviolet etc.) 
are neglected in this study 
since we treat protons above $10^{19.6}$eV. 
Such protons remain nearly unaffected by the higher energy background photons 
because of their relatively small number. 
Protons above $8 \times 10^{19}$eV lose their energies 
by photopion production, $p + \gamma \longrightarrow \pi + X$. 
This reaction has large inelasticity ($\sim$ 30\%) 
and relatively short energy-loss length of about a few ten Mpc at $z=0$. 
Protons with such energies cannot reach the Earth from distant sources. 
Thus, the photopion production predicts sharply spectral steepening 
around $8 \times 10^{19}$eV, 
so-called Greisen-Zatsepin-Kuz'min (GZK) steepening 
\citep*{greisen66,zatsepin66}. 
The GZK effect is essential in considering the correlation 
with distribution of relatively distant sources. 
We adopt the energy-loss length 
which is calculated by simulating the photopion production 
with the event generator SOPHIA \citep*{mucke00}. 
On the other hand, 
protons below $8 \times 10^{19}$eV lose their energies 
mainly due to Bethe-Heitler pair creation, 
$p + \gamma \longrightarrow p + e^+ + e^-$. 
This has small inelasticity ($\sim 10^{-3}$). 
We adopt the analytical fit function 
given by \cite{chodorowski92} to calculate 
the energy-loss rate on isotropic photons. 
An adiabatic energy loss due to the expanding universe 
also makes protons lose their energies. 
The energy-loss rate is written as 
\begin{equation}
\frac{d E}{dt} = -\frac{\dot{a}}{a}E = 
-H_0 \left[ \Omega_m (1 + z)^3 + \Omega_{\Lambda} \right]^{1/2} E. 
\end{equation}
These three energy-loss processes are treated as continuous processes 
in our calculation. 

Trajectories of protons are deflected by EGMF. 
Protons injected from their sources to the Earth cannot be led into the Earth 
straightforward. 
It wastes much CPU time to calculate the propagation of protons 
which cannot arrive at the Earth in order to construct the arrival distribution. 
For solving such problem, 
we suggested a new calculation method of the arrival distribution 
which is an application of the backtracking propagation \citep*{takami06}. 
In this method, 
protons with the charge of -1 are ejected from the Earth 
and then we calculate their trajectories in intergalactic space 
taking into account magnetic deflections and energy-loss(gain) processes. 
Such trajectories are regarded as those of protons from extragalactic space. 
We calculate only trajectories of protons which can reach the Earth. 

In order to simulate the arrival distribution, 
2,500,000 protons (with charges of -1) with 
$dN/d\log_{10}E \propto {\rm Const.}$ from $10^{19}$ to $10^{21}$eV. 
are ejected isotropically from the Earth. 
We calculate their trajectories until their propagation time exceeds 
the lifetime of Universe or their energies reach $10^{22}$eV, 
which corresponds to the maximum acceleration energy at UHECR sources. 
For each source distribution, 
we calculate a factor for a trajectory of $j$ th particle, 
which represents the relative probability that the $j$ th proton 
reaches the Earth, 
\begin{equation}
P_{\rm selec}(E,j) \propto \sum_i 
\frac{1}{( 1 + z_{i,j}) {d_{i,j}}^2} 
\frac{dN / dE_g (d_{i,j}, {E_g}^{i})}{E^{-1.0}} 
\frac{d E_g}{dE}. 
\end{equation}
Here $i$ labels sources on each trajectory, 
while $z_{i,j}$ and $d_{i,j}$, are 
redshift and distance of such sources. 
${E_g}^i$ is the energy of a proton at the $i$ th source, 
which has the energy $E$ at the Earth. 
Thus, $dN / dE_g (d_{i,j}, {E_g}^i) \propto {E_g}^{-2.6}$ 
is the energy spectrum of UHE protons ejected from a source 
whose distance is $d_{i,j}$. 
$d E_g / dE$ represents correction factor for the variation of the shape of 
the energy spectrum through the propagation. 

We randomly select trajectories according to these relative
probabilities, ${\rm P}_{\rm selec}$,  
so that the number of the selected trajectories 
is equal to the required event number. 
The mapping of the ejected direction of each particle at the Earth 
is the arrival distribution of UHE protons. 
If we have to select the same trajectory more than once 
to adjust the number of selected trajectories, 
we generate new events 
whose arrival direction is calculated by adding a normally 
distributed deviation with zero mean and 
variance equal to the experimental resolution 
to the original arrival direction. 
The experimental resolution is assumed to be $1^{\circ}$. 

\begin{figure*}
\begin{center}
\epsscale{2.0}
\plotone{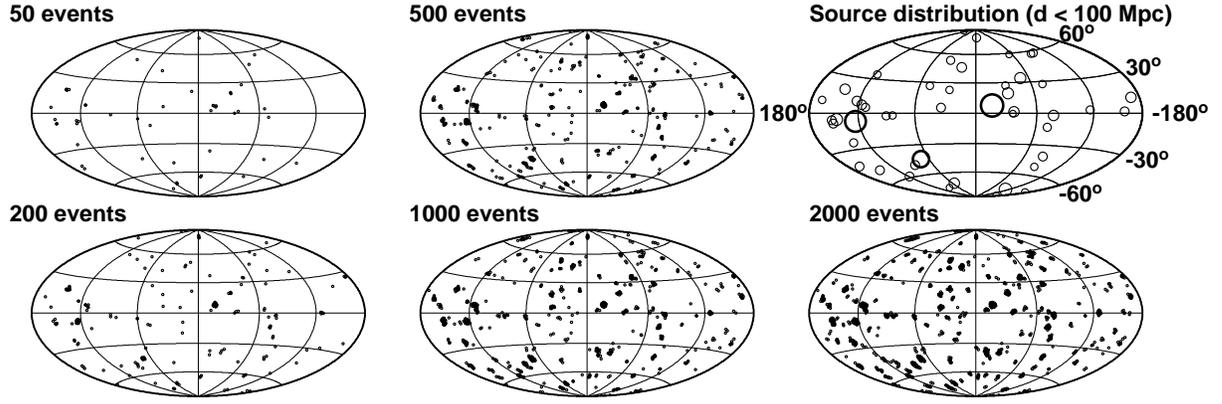}
\caption{The arrival distributions of UHE protons above $10^{19.8}$eV 
predicted by a specific source distribution with $10^{-5}~{\rm Mpc}^{-3}$ 
({\it upper right}) in the galactic coordinate. 
The EGMF is not included. 
The source distribution within 100 Mpc is shown 
as radii of circles inversely proportional to source distances. 
The sources within 50 Mpc are shown with bold circles. 
The simulated number of events are 50 ({\it upper left}), 
200 ({\it lower left}), 500 ({\it upper middle}), 1000 ({\it lower middle}), 
and 2000 ({\it lower right}).}
\label{fig:corF5000G0E0N47L0}
\end{center}
\end{figure*}

\subsection{Statistical Quantities} \label{method_statistics}

In order to investigate statistically the similarity 
between the arrival distribution of UHECRs $f_e$ and 
the source distribution $f_s$, 
a correlation coefficient between the two distributions is defined as 
\begin{equation}
\Xi(f_e,f_s) \equiv \frac{\rho(f_e,f_s)}
{\sqrt{\rho(f_e,f_e) \rho(f_s,f_s)}}
\label{eq_cc}
\end{equation}
where
\begin{equation}
\rho(f_a,f_b) \equiv \sum_{j,k} 
\frac{f_a(j,k)-\bar{f_a}}{\bar{f_a}} 
\frac{f_b(j,k)-\bar{f_b}}{\bar{f_b}} 
\frac{\Delta\Omega(j,k)}{4\pi}. 
\end{equation}
Here subscripts $j$ and $k$ distinguish each cell of the sky, 
$\Delta\Omega(j,k)$ denotes the solid angle of the $(j,k)$ cell, 
and $\bar{f_a}$ means the average of $f$ calculated as 
\begin{equation}
\bar{f_a} \equiv \sum_{j,k} f_a(j,k) \frac{\Delta\Omega(j,k)}{4\pi}. 
\end{equation}
By definition, 
$\Xi$ ranges from -1 to +1. 
When $\Xi=+1(-1)$, the two distributions are exactly the same(opposite). 
When $\Xi=0$, we cannot find any resemblance between the two distributions. 
A source at a distance $d_i$ approximately contributes to arrival cosmic rays 
with the weight of ${d_i}^{-2}$ 
since all sources are assumed to have the same injection power. 
Therefore, $f_s(j,k)$ is calculated as $\sum_i 1/{{d_i}^2}$, 
where $i$ runs over sources in the $(j,k)$ cell. 

\section{Results} \label{results}

For a given source distribution, 
we can simulate the arrival distribution of UHE protons 
using the calculation method explained in the previous section. 
We investigate the correlation between the arrival distribution 
and the source distribution, 
and discuss the possibility to unveil the local structure by UHECRs. 

\begin{figure*}
\begin{center}
\epsscale{1.35}
\plotone{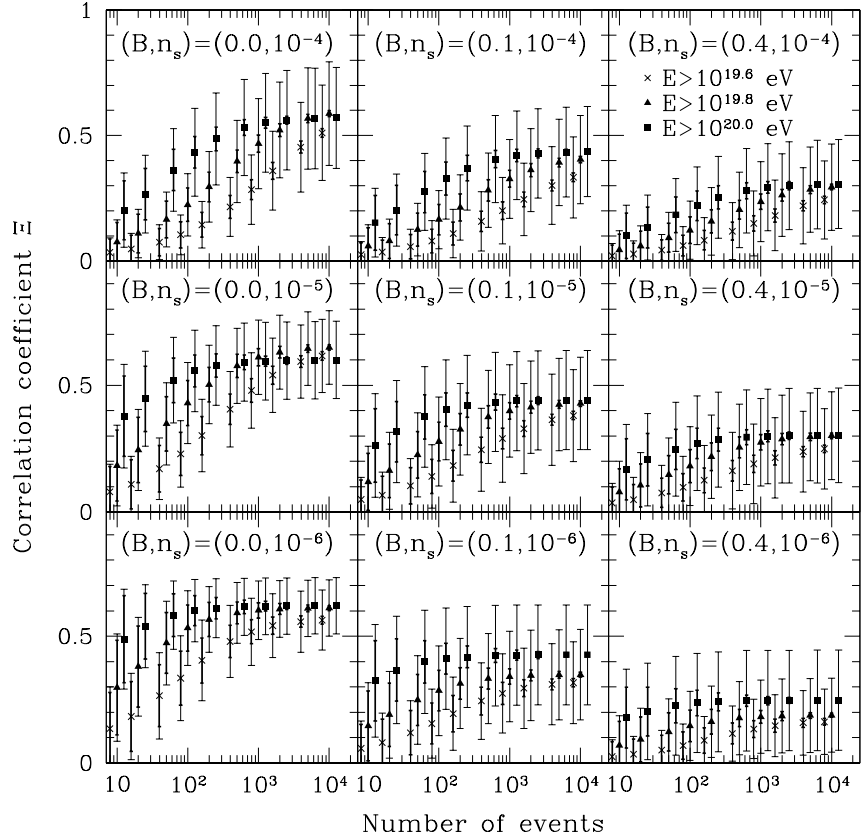}
\caption{Correlation coefficients 
between the arrival distribution of UHE protons 
above $10^{19.6}$ ({\it crosses}), $10^{19.8}$ ({\it triangles}) 
and $10^{20.0}$eV ({\it squares}),  
and their source distribution within 100 Mpc 
for several EGMF strengths and the number densities of UHECR sources 
as a function of expected number of cosmic rays on all the sky. 
The size of the cell for the calculation of the correlation coefficients 
is chosen to be $2^{\circ} \times 2^{\circ}$. 
The three marks on the same number of events are a little shifted 
horizontally for visibility. 
Units of $B$, normalization factor of the EGMF strength, 
and $n_s$, the source number density, is $\mu$G and Mpc$^{-3}$ respectively. 
The short (thick) error bars represent the fluctuations 
due to the finite number of observed events 
and the long (thin) error bars are total errors 
including the cosmic variance.}
\label{fig:corD100G0R2}
\end{center}
\end{figure*}

In Fig.\ref{fig:corF5000G0E0N47L0}, 
we demonstrate arrival distributions of UHE protons above $10^{19.8}$eV 
predicted by a specific source distribution 
with $10^{-5}~{\rm Mpc}^{-3}$ in the galactic coordinate. 
The EGMF is not included. 
The source distribution within 100 Mpc is shown on its upper right panel. 
The radii of circles in this figure are inversely proportional to 
source distances. 
The sources within 50 Mpc are shown with bold circles. 
The number of simulated events are 50 ({\it upper left}), 
200 ({\it lower left}), 500 ({\it upper middle}), 1000 ({\it lower middle}), 
and 2000 ({\it lower right}). 
Note that 50 events on all the sky correspond to the Auger result at present 
since Auger observes cosmic rays in the southern hemisphere \citep*{roth07}. 

When 200 events are detected, 
strong event clusterings from nearby sources (within 50 Mpc) can be observed. 
The distribution of more distant sources are not found yet 
except for the directions that sources are positionally concentrated. 
Detection of more than 500 events enables us to find event clusterings 
in the direction of almost all sources within 100 Mpc. 
Thus, we can find graphically 
that detection of about 500 cosmic rays above $10^{19.8}$eV 
makes us unveil nearby source distribution of them. 

We should discuss statistically (not graphically) 
the probability that future observations 
can unveil UHECR source distribution. 
Fig.\ref{fig:corD100G0R2} shows predicted correlation coefficients 
between the arrival distribution of UHE protons 
and their source distribution within 100 Mpc 
for several EGMF strengths and source number densities 
as a function of expected number of cosmic ray events. 
The size of the cell for the calculation of the correlation coefficients 
is chosen to be $2^{\circ} \times 2^{\circ}$ 
since angular resolution of UHECR experiments, $1^{\circ}$, is 
taken into account in simulating the arrival directions. 
The numbers of events that we calculate the correlation coefficients 
are 10, 20, 50, 100, 200, 500, 1000, 2000, 5000, 10000 on all the sky 
for $E > 10^{19.6}$ ({\it crosses}), $E > 10^{19.8}$ ({\it triangles}), 
and $E > 10^{20.0}$eV ({\it squares}). 
The three marks on the same number of events are slightly shifted 
horizontally for visibility. 
The EGMF strengths, $B$,  and the source number densities, $n_s$,  
which are adopted for the simulation are represented on each panel. 
There are two errors. 
The thick (short) error bars show the statistical error, i.e., 
the fluctuations due to the finite number of observed events, 
averaged over 100 realizations of the source distribution 
for a given source number density. 
The random event selection to estimate the fluctuations 
is performed 100 times for each source distribution. 
On the other hand, the thin (long) error bars represent the total error, 
i.e., the statistical error plus the cosmic variance. 
The latter error bars also include 
the variation between different realizations of source distributions. 
Note that statistical errors on an observation are 
estimated as the thick error bars 
since only a source distribution is true in the Universe. 

\begin{figure*}
\begin{center}
\epsscale{2.0}
\plotone{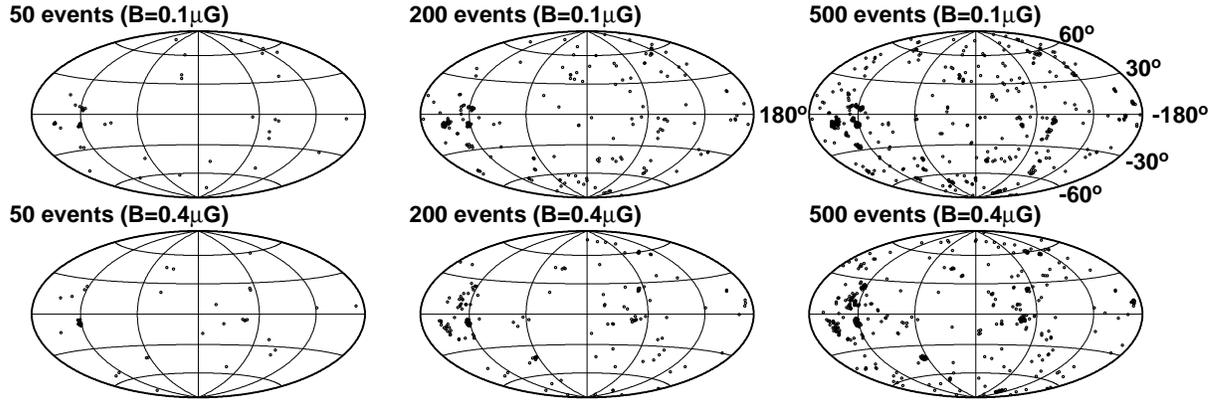}
\caption{The arrival distributions of UHE protons above $10^{19.8}$eV 
predicted by the same source distribution used in Fig.2. 
The upper panels show those taking into account propagation 
in the EGMF with $B=0.1\mu$G 
and the lower panels are the same, but $B=0.4\mu$G.}
\label{fig:corF5000G0E14N47L0}
\end{center}
\end{figure*}

First, we deliberate the correlation in the case of no magnetic field. 
A lot of features of the correlation coefficients common to finite EGMF cases 
are found in this simplest case. 
The three left panels show the correlation coefficients 
in the case of no EGMF. 
When the numbers of observed events increase, 
the correlation coefficients start to converge, 
and the final values can be estimated. 
The statistical errors (thick error bars) decrease. 
The cosmic variance remains at large number of events 
since it is uncertainty to originate from different source distributions. 
Note that a source distribution used in Fig.\ref{fig:corF5000G0E0N47L0} 
predicts correlation coefficients 
similar to the averages of those in the middle left panel 
of Fig.\ref{fig:corD100G0R2}. 
The correlation coefficient cannot converge on 1.0 
in spite of rectilinear propagation 
due to the finite angular resolution, the energy loss processes, 
and also because cosmic rays do not always come from sources within 100 Mpc. 
Since UHE protons from more distant source needs 
the higher energy at the source in order to reach the Earth 
with the same energies, 
such source contributes to arrival cosmic rays smaller than nearer source 
even if a factor of ${d_i}^{-2}$ is taken into account. 
In other words, the arrival distribution of many cosmic rays 
is only approximately the same spatial pattern 
as their source distribution weighted by ${d_i}^{-2}$. 

\begin{figure*}
\begin{center}
\epsscale{1.35}
\plotone{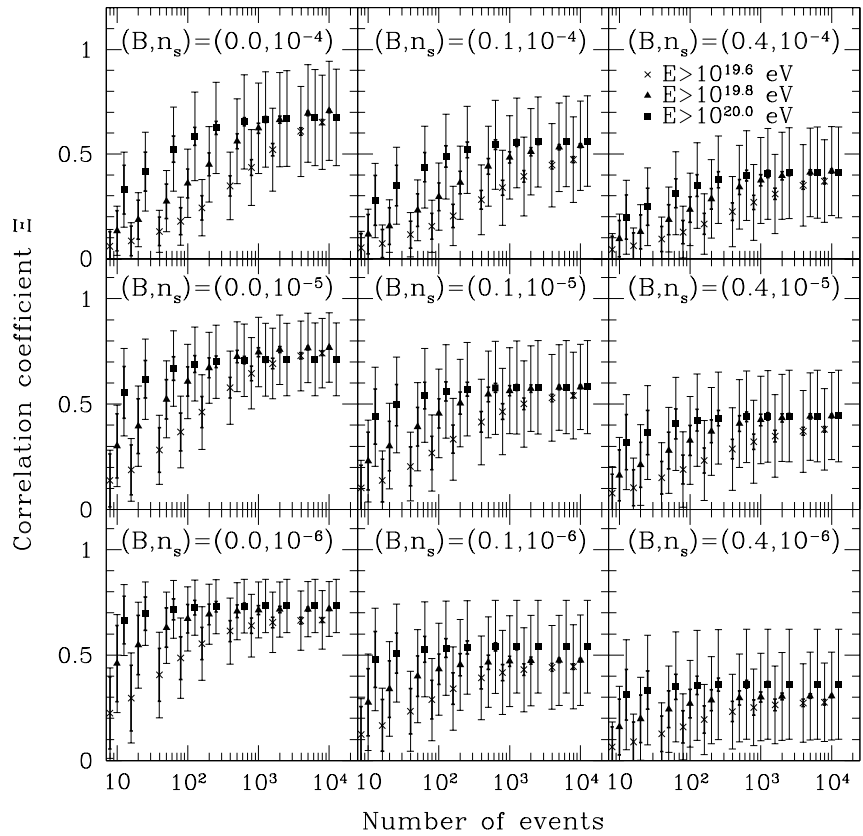}
\caption{Same as Fig.\ref{fig:corD100G0R2}, 
but for $4^{\circ} \times 4^{\circ}$.}
\label{fig:corD100G0R4}
\end{center}
\end{figure*}

We focus on the behaviors of averages of the correlation coefficients 
on each panel. 
The correlation coefficients with protons above $10^{20}$eV predict 
larger values than those with lower energies 
at the small number of observed events. 
As the number of detected events increases, 
the coefficients with highest energies 
converge on the smaller values 
than those with lower energies. 
The reason is GZK mechanism. 
UHE protons above $10^{20}$eV can come from sources 
only within the GZK sphere ($\sim 50$Mpc). 
On the other hand, 
protons with lower energies can arrive at the Earth outside 100 Mpc. 
Thus, the correlations are better at the small number of events. 
The difference between the correlation coefficients 
with protons above $4 \times 10^{19}$eV and those above $10^{20}$eV 
is statistically significant at 1$\sigma$ level 
since strengths of errors on observation 
are estimated as the thick error bars. 
Since the radius of the GZK sphere is less than 100 Mpc, 
there are sources within 100 Mpc which cannot be traced 
by cosmic rays above $10^{20}$eV. 
Thus, as the number of observed events increases, 
the correlation coefficients with such highest energy protons 
converge on a little smaller values than those with lower energy protons. 
The correlations of protons with $E > 10^{19.8}$eV 
are better than those with $E > 10^{19.6}$eV, 
because lower energy protons can reach the Earth from more distant sources. 

Next, we investigate the correlation including EGMF. 
The three middle panels and the three right panels 
of Fig.\ref{fig:corD100G0R2} show the correlation coefficients 
in the case of $B=0.1$ and $0.4\mu$G respectively. 
The EGMF diffuses trajectories of UHECRs during their propagation 
and obscures their arrival directions \citep*{takami06}. 
Therefore, stronger EGMF predicts weaker correlation 
in any source number density. 
Compared to the correlation coefficient in the case of no EGMF, 
that of protons with the lower energy threshold is worse 
since lower energy protons are deflected by EGMF more largely. 
Protons with $E > 10^{19.6},~10^{19.8}$eV predicts 
smaller correlation than those above $10^{20}$eV.

The number of events that the correlation coefficients start to converge 
is weakly dependent on the EGMF strength 
and depends on the source number density. 
In the case of $10^{-6}~{\rm Mpc}^{-3}$, 
such numbers of events are about 1,000, 200, and 50 
for $E > 10^{19.6},~10^{19.8}$ and $10^{20}$eV respectively. 
At larger number densities, such numbers increase. 
In the case of $10^{-5}~(10^{-4})~{\rm Mpc}^{-3}$, 
these  numbers are about 5,000($>$10000), 500(5000), and 100(500) 
for $E > 10^{19.6},~10^{19.8}$ and $10^{20}$eV respectively. 
Detection of such number of events enables us to unveil 
UHECR source distribution at visible level by UHE protons, 
which concerned with the final value of the correlation coefficients. 
Both cosmic rays above $10^{19.8}$eV and those above $10^{20}$eV are 
good indicators for unravelling the source distribution 
since their correlation values converge to nearly equal values 
for $10^{-4}$ and $10^{-5}~{\rm Mpc}^{-3}$. 
However, the detected number of cosmic rays above $10^{19.8}$eV 
reaches the number that the correlation coefficient starts to converge 
in shorter observing period than that of cosmic rays above $10^{20}$eV 
since cosmic rays above $10^{20}$eV are almost not detected 
in the presence of the GZK steepening. 
Note that our source model predicts the GZK steepening. 
Thus, UHE protons above $10^{19.8}$eV are best indicators 
to unveil the source distribution within 100 Mpc. 
Note that ground based detectors such as Auger and Telescope Array (TA) 
\citep*{ta} can observe only about half hemisphere, 
so only half of such numbers are observed. 

Fig.\ref{fig:corF5000G0E14N47L0} is the simulation of 
the arrival distribution of UHE protons above $10^{19.8}$eV 
in the case of $B=0.1$ and $0.4\mu$G. 
The source distribution is the same one in Fig.\ref{fig:corF5000G0E0N47L0}. 
In Fig.\ref{fig:corF5000G0E0N47L0} and \ref{fig:corF5000G0E14N47L0}, 
the arrival distributions with 200 protons 
(less than 500 events, which is the number 
that the correlation coefficients start to converge) 
include several number of event clusterings. 
Compared to the three cases of the EGMF strength, 
the arrival directions of protons are diffused by EGMF. 
We can also find event clusterings in the directions of 
nearby sources or source clusterings. 
Such event clusterings are not lost by a structured EGMF 
although their angular scales are spread. 
At 500 event detection, 
the event clusterings increase and 
the local structure of the universe appears.

The larger size of the cell for the calculation of 
the correlation coefficients is expected to 
lead to larger correlation coefficients 
since EGMF obscures UHECR arrival directions as mentioned above. 
Fig.\ref{fig:corD100G0R4} shows the same figure, but 
the size of the cell is chosen to be $4^{\circ} \times 4^{\circ}$. 
The correlation coefficients are larger 
than those in Fig.\ref{fig:corD100G0R2} on all panels. 
These increase even in the case of $B=0.0\mu$G 
since a part of arrival cosmic rays from a source is distributed 
in a small scale, but larger than $2^{\circ}$ 
due to finite angular resolution of UHECR experiments. 
The values of the coefficients are slightly not increase 
even if the size of the cell is larger than $4^{\circ} \times 4^{\circ}$. 
In this figure, we find that the numbers of events starting 
the convergence of the correlation coefficients are almost unchanged. 

\section{Conclusion} \label{conclusion}

In this paper, 
we calculated the arrival distribution of UHE protons 
taking into account energy losses and deflections by EGMF 
during propagation in intergalactic space 
in order to investigate the possibility that 
future observations of UHECRs can unveil the local structure of UHE universe. 
In order to reproduce a realistic situation, 
we adopted a structured EGMF model and source distributions 
which reproduce the observed local structures. 
The arrival distribution of UHE protons was compared statistically to 
their source distribution using the correlation coefficients. 
As the number of observed events increases, 
the correlation coefficient increases 
and converges to some value 
which represents the ability to unveil 
the source distribution by UHE protons, i.e. charged particles. 
Thus, the number of events that the correlation coefficient 
starts to converge is an important number for UHECR observations. 
In other words, 
detection of such number of events allows us to unravel 
UHECR source distribution. 
We found that UHECRs above $10^{19.8}$eV are best indicators 
to decipher their source distribution within 100 Mpc 
from discussion based on the final values of the correlation coefficients 
and GZK mechanism, 
and 5000, 500, and 200 event detections above $10^{19.8}$eV on all the sky 
can unveil their source distribution for the source number densities of 
$10^{-4}$, $10^{-5}$, and $10^{-6}~{\rm Mpc}^{-3}$ respectively. 
Note that ground based detectors observe only about half hemisphere, 
so only half of such numbers are requested. 

In this study, 
we took only EGMF into account as magnetic field, i.e., neglected GMF 
GMF deflects trajectories of UHE protons efficiently 
by its regular components, 
which consist in spiral 
and dipole components \citep*{alvarez02,yoshiguchi03b}. 
A turbulent component of GMF very weakly change 
the arrival directions of UHE protons \citep*{yoshiguchi04}. 
The deflection angles of UHECR protons are a few degree 
at around $10^{20}$eV except for the direction of Galactic Center. 
Such deflection disturbs the spatial pattern of UHECR arrival distribution 
at a few degree scale. 
The effect of GMF is one of our future investigations. 

A lot of inquiries on UHECR source number density result in 
around $10^{-5}~{\rm Mpc}^{-3}$ on ground of the AGASA results 
as introduced in section \ref{intro}. 
If this number density is true, 
five year observation by Auger and 
future observation by TA and Extreme Universe Space Observatory \citep*{euso}
will reveal the distribution of nearby UHECR sources. 
The dawn of the UHE particle astronomy is just around the corner. 

\acknowledgments

The work of H.T. is supported by Grants-in-Aid for JSPS Fellows. 
The work of K.S. is supported by Grants-in-Aid for 
Scientific Research provided by the Ministry of Education, 
Science and Culture of Japan through Research Grants S19104006.

\end{document}